# A Taxonomy of Modeling Approaches for Systems-of-Systems Dynamic Architectures: Overview and Prospects

Ahmad Mohsin [*1,2,3], Naeem Khalid Janjua, Syed MS Islam, Valdemar Vicente Graciano Neto [4]

[*1,2,3] School of Science, Edith Cowan University, Joondalup, WA, Australia,{a.mohsin, n.janjua, syed.islam}@ecu.edu.au
[4] Instituto de Informática (INF), Universidade Federal de Gois (UFG) Goiânia, GO, Brazil, valdemarneto@inf.ufg.br

*Abstract*—Systems-of-Systems (SoS) result from the collaboration of independent Constituent Systems (CSs) to achieve particular missions. CSs are not totally known at design time, and may also leave or join SoS at runtime, which turns the SoS architecture to be inherently dynamic, forming new architectural configurations and impacting the overall system quality attributes (i.e. performance, security, and reliability). Therefore, it is vital to model and evaluate the impact of these stochastic architectural changes on SoS properties at abstract level at the early stage in order to analyze and select appropriate architectural design. Architectural description languages (ADL) have been proposed and used to deal with SoS dynamic architectures. However, we still envision gaps to be bridged and challenges to be addressed in the forthcoming years. This paper presents a broad discussion on the state-of-the-art notations to model and analyze SoS dynamic architectures. The main contribution this paper is threefold: (i) providing results of a literature review on the support of available architecture modeling approaches for SoS and an analysis of their semantic extension to support specification of SoS dynamic architectures, and (ii) a corresponding taxonomy for modeling SoS obtained as a result of the literature review. Besides, we also discuss future directions and challenges to be overcome in the forthcoming years.

## I. INTRODUCTION

The recent advancements in inter-connectivity, computing technologies and integration of heterogeneous systems has given birth to a new type of software-intensive systems called Systems-of-Systems (SoS). The implementation of these systems in various domains is increasing, demanding high quality in their overall design and development. SoS are complex systems resulting from the interoperability of Constituent Systems (CS), managing resources and capabilities with managerial and operational independence that collaborate to produce emergent behaviors to achieve a specified global mission [1]. A remarkable example of SoS is smart city, which is composed of many individual systems that manage the life in a city, such as monitoring emergencies, managing power distribution, and controlling the traffic [2], [3]. Since these systems play a vital role in human lives and critical infrastructure, therefore they require rigorous, architectural modeling approaches [4], to specify and reason their structure (CSs, connectors and underlying properties), behavior (the way they interact) and configurations (temporary alliance resulting from CSs interactions) [5], [6].

In a typical SoS, CSs are autonomous, which turns the SoS architecture to be inherently dynamic. SoS dynamic architecture makes the SoS assume several different architectural configurations (also known as *coalitions*), which brings (i) a non-deterministic nature to it due to addition, deletion and updates in CSs at runtime, and (ii) high degree of unpredictability, which potentially impact SoS functional and non- functional properties [7]. Hence, despite being a complex and difficult task, modeling these stochastic dynamic architectural changes early in the life cycle at SoS abstract level becomes a prominent endeavour.

Traditionally, Architectural Description Languages (ADLs) are based on strong mathematical foundations of process algebras [8] and have been used for modeling large but mostly stand-alone single systems. In this regard, there are a number of studies [9], [10] reporting various methods for modeling static and dynamic architectures of centrally developed single systems, with formal ADLs, including Darwin, Rapide, Unicon and Wright [11]–[13]. Conversely, there are only few studies that describe approaches to model SoS dynamic architectures. And, among those proposals, most of them fail to provide a broader view on various architectural aspects such as dynamic structural changes, formation of new configurations, and quantitative analysis of these changes on the overall SoS architecture [13]–[15].

The main contribution of this paper is then providing foundations for forthcoming research on SoS dynamic architectures. In this regard, we provide results of literature review, and a respective taxonomy of ADLs and related techniques for SoS, discussing their semantic potential to represent SoS dynamic architectures and support the prediction of coalitions and impact of changes on behaviors and non-functional properties of that SoS.

The remaining of the paper is organized as follows: Section II describes SoS architectural concepts, types and its dynamic nature. Section III presents the proposed taxonomy for ADL for describing SoS, besides the results of the literature review discussed under the taxonomy identified classes. Section IV raises the research issues identified during the investigation of ADL for SoS and proposition of the taxonomy. Section V presents future research directions and finally, conclusions are provided in section VI.

## II. SoS ARCHITECTURAL CONCEPTS, TYPES AND DYNAMICTY

SoS are complex type of systems with key architectural elements such as CSs, mediators (connectors for CSs) and coalitions (resulting configurations due to CSs interactions). Since, their associated architectural concepts are different from traditional single systems essentially due to unique characteristics of independence, emergence, evolution, geographic distribution, and dynamicity, therefore, it is useful to define such concepts (refer to Table 1) that are used in SoS modeling.

TABLE I
SoS-ARCHITECTURAL MODELING CONCEPTS

| SoS-Architectural Concepts | Description |
|---|---|
| Structural Description | A structural SoS description includes CSs, mediators, and interfaces and overall relationships among them with constraints and properties at abstract level. |
| Behavioral Description | SoS behavioral description specifies and reasons about CSs actions/activities they participate to produce emergent behaviors. This also includes specification of mediator to mediate a particular interaction between CSs. |
| Dynamic Architecture | SoS dynamic architecture is concerned with change in configurations among interacting CSs and mediators at runtime leading to dynamic emergent behavior of the SoS during a single computation. |
| Dynamic Reconfiguration | It is concerned about describing reconfigurations due to dynamic architectural changes in SoS structure such as addition, deletion of CSs and connectors at runtime with unplanned interventions. SoS dynamic reconfigurations are stochastic in nature. |

Correct and consistent software plays a vital role towards successful engineering of an SoS, motivating research community to establish strategies, techniques, and standards to deal with the modeling concerns to engineer high-quality SoS [8]. Since SoS are software-intensive systems, these systems essentially exhibit a software architecture, which composes the SoS in its fundamental structure, with architectural elements including its CSs, connectors, underlying properties of the these elements and of the environment [5]. Due to the key characteristics of operational and managerial independence of constituents [1], SoS software architectures are inherently dynamic, since constituents can freely join or leave the SoS structure at a certain time leading to new configurations [6]. Such ability can be considered even an essential advantage, as it possibly minimizes system disruptions while new or modified constituents are joined into a SoS to substitute the failing ones. However, such characteristic also increases the level of uncertainty about SoS operation, and its conformance to overall system properties.

To realize the impact of complexity and underlying managerial and operational independence of CSs on SoS configurations, it is important to understand SoS classification and their impact on architectural modeling. In this regard, Maiers classification [1] is mainly based on management policies and governance are defined for SoS. Dahmann et al. [16] extended Maier classification and introduced Acknowledge

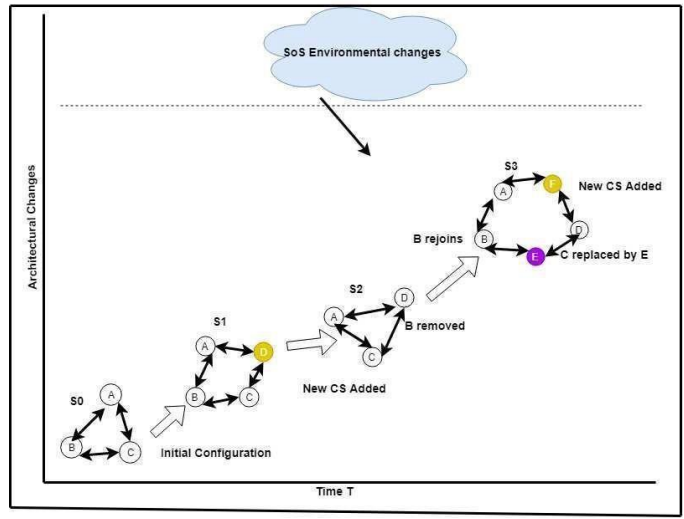

Fig. 1. SoS architectural changes and states formations

SoS emphasizing collaborative management at global level. They are formally described as follows:

**Directed SoS:** The CSs are independent but managed centrally by a single authority with common policies to achieve SoS objectives.

**Collaborative SoS:** There is no central authority in this type of SoS instead CSs are independent to operate and managed. Resulting configurations are formed with agreed policies but are subject to individual CSs participation.

**Acknowledged SoS:** Acknowledged SoS are formed with recognized objectives, resources, and designated manager but CSs maintain their operations and to some extent managerial independence.

**Virtual SoS:** This type of SoS has no central controlling authority and lack common objectives, CSs may join or leave at any time, leading to high emergence and volatile system structure.

During the modeling of SoS, the decision to use a particular type of SoS impacts on the overall control that a team has on specifying SoS architecture since each type provides different level of managerial and operational independence to CSs.

SoS are complex stochastic systems exhibiting random and dynamic structural changes while randomness in SoS states makes it a stochastic process leading to uncertainty. Dynamicity and stochasticity are inter-related, therefore, SoS experiences random configurations due to runtime changes according to the following operations [17]:

1) **Addition**: A new CS is added to the SoS,
2) **Removal**: CS may be removed from the system,
3) **Replacement**: CS may be replaced from the system, i.e., removed and substituted by a new similar one, and
4) **Rearrangement**: the complete architecture can be dissolved and rebuilt in a different arrangement.

Figure 1 depicts the scenario from initial configurations of a typical SoS to new configurations with runtime perspective involving CSs (A,B,C D, E, and F) leading to unplanned reconfigurations. Each configuration presents a particular state 's' of SoS architecture at a particular time 't' where 'T' is overall timeline. Every dynamic reconfiguration state emerges with its own quality attributes that may deviate from the stated requirements. Apart from these changes, there are some other factors that impact SoS structure, such as CSs internal changes and SoS environmental level changes [18].

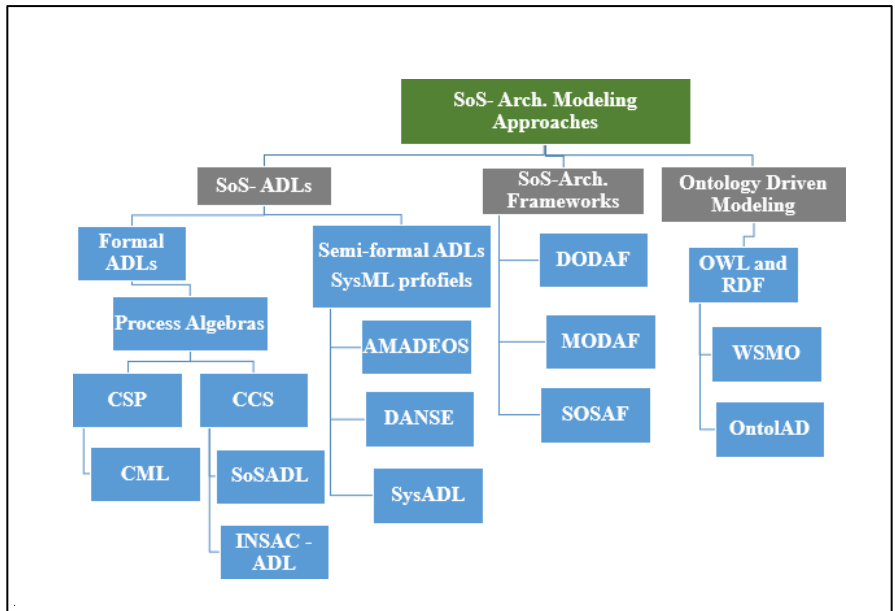

Fig. 2. A taxonomy for SoS architecture modeling approaches.

### III. LITERATURE REVIEW AND A TAXONOMY OF ADL FOR SOS

Although modern architectural description languages (ADL) support both representation and validation of properties of SoS software architectures, they lack of mechanisms to aid the validation of dynamic features of SoS. A literature review was performed to obtain a panorama of the existing approaches that support modeling, reasoning, and analysis of SoS dynamicity. From this review, a taxonomy has been developed to categorize these approaches according to their ability to model SoS, as depicted in Figure 2. This taxonomy is categorized into SoS-ADLs, SoS architecture frameworks, and ontology-driven SoS modeling. These techniques have been evaluated on the basis of structural, behavioral, dynamic reconfiguration and quality attributes modeling, reasoning abilities along with model-driven analysis support of individual approaches for SoS where these architectural concepts have already been defined in Table 1. In the following sections, we discuss the results obtained in our literature review using the classes identified in the taxonomy.

#### A. SoS-ADLs

There are different ADLs for SoS modeling however, their ability to express system requirements into architectural view is still an issue [19]. Nielsen et al. [5] describes the main concerns that must be taken care by an ADL are : (i) well-founded interfaces (ii) description of autonomous behaviors, capabilities, and responsibilities of CSs, (iii) description of policies for independently modeling CSs and (iv) description of desired and undesired behaviors, observable by abstract models. However, according to Malalvolta et al. [20], ideally there is no single ADL to possess all these characteristics for describing and modeling architecture. The underlying formalism on which an ADL is based is critical to assess an ADL for its features among other characteristics. For the this we have categorized SoS- ADLs into semi-formal ADLs and formal ADLs as follows:

*1) Semi-Formal (SysML Profiles):* Starting from semi-formal languages, SysML[1] is a widely used ADL for modeling and analyzing SoS architecture by employing Model Driven Engineering (MDE) [21]. Researchers and practitioners have used it from different perspectives when it comes to modeling SoS. Some have only used it to model structural elements of SoS while others have tried to build strong profiles of SySML by employing OCL and temporal logic to model and analyze SoS complex behavior. DANSE[2], AMADEOS[3], and COMPASS[4] are most recent projects where SysML has been used as the main language to deal with different architectural aspects of SoS.

In AMADEOS SoS dynamicity, emergence and evolution have been modeled using SysML profiles [22]. Various architectural views have been presented with different profiles to describe structure, behavior and dynamic models. However, the problem of dynamic reconfiguration is not achieved as SysML profiles do not have the ability to specify CSs interfaces to SoS environment when runtime changes occur. Authors in DANSE project try to formalize SoS modeling using SysML/UPDM profiles [23] where CSs are distinguished with their participating roles, named as capabilities with the strong specification of interface contracts to take part in a particular mission. It further extends CSL (Contracts Specification Language) to formulate formal semantics to guarantee CSs behavior with its integration to SoS.

---
[1]https://www.omg.org/spec/SysML/
[2]https://danse-ip.eu/home/model-sos.html
[3]http://amadeos-project.eu/
[4]http://www.compass-research.eu/

UML behavioral modeling with OCL and SysML extensions [24] are used to describe various behaviors especially to improve loose coupling for interfaces of constituents, where contracts of interaction have been defined. However, these are not capable to model reconfigurations or to reason about dynamic architecture as underlying semantics are essentially semiformal. Recently SysADL [25] has been presented as an ADL with SysML profiles to describe structural, behavioral and executable views of SoS architectures. However, it has certain limitations when it comes to model and reason reconfigurations and explicit representation of quality attributes. Table 2 provides a detailed critical analysis of SysML based semi-formal approaches for modeling SoS.

*2) Formal SoS - ADLs:* Most of the existing formal SoS ADLs are based on process algebraic formalism of Communicating Sequential Processes (CSP) [26] and Calculus of Communicating Systems (CCS) [27] by employing MDE. SoS are complex dynamic systems for which formal ADLs based on strong mathematical foundations are suitable to model and analyze architecture for effective design of the system. Analysis of formal SoS-ADLs are summarized in Table 3.

Compass Modeling Language (CML) [28] is a formal modeling language for SoS, based on CSP and VDM (Vienna Development Method) [26], [29] to model, reason and simulate behavioral architecture with a focus on fault tolerance and dependability analysis. CML uses processes to present CSs where channels are used for communication. It allows CSs to explicitly define their interface contracts to SoS environment and to further integrate with SysML for the better semantics. But since CML is a low-level formal language where SysML is a semi-formal language the corresponding mapping between the two creates additional complexities.

Archware started a series of industrial projects through which they have been able to develop a formal ADL for SoS named as SosADL [30]. This language is based on formal foundations of SoS Pi-Calculus [31] originated from CCS and CCP [32] to support the architecture-centric modeling and analysis while describing constrained behaviors to deal with emergent behavior in dynamic environments of SoS. In SosADL concrete architectures are generated at runtime from abstract architecture to deal with dynamicity. Endogenous reconfigurations are simulated using Discrete Event System Specification (DEVS) from abstract architecture defined in SosADL using model transformations [17]. However, the mechanism for predicating configurations at SosADL level is still lacking. The simulation of reconfiguration is performed with DEVS formalism using Dynamic Reconfiguration Controller (DRC) adding overall complexity to the process. Moreover, this technique is unable to model and analyze possible reconfigurations at runtime and their impact on quality attributes. As a formal approach reaction rules have been used for rewriting bigraphs [33] for modeling and simulation of SoS. This technique allows to represent SoS architectural elements in bigraphs as agents to model the architecture and specifies structure and behavior with reactive rules. Dynamic behavioral modeling using this technique is still infancy.

Based on Pi-calculus INSAC-ADL [34] is a formal approach specifically designed to model CPS-SoS. It captures meta-models of structural and dynamic architecture using Goal-Oriented requirements engineering technique. For runtime interactions of elements, authors have used IPL (INSAC prospection language) while for dynamics they used IML (INSAC modeling language). Here INSAC utilizes both IPL and IML for architecture modeling. Since it is based on formal foundations of Pi-calculus it is able to describe static and dynamic architecture but lacks the capability to build and expresses strong vocabulary to be able to reason about unforeseen collations (configurations) among constituent systems. Archsos is a relatively new ADL [35] specifically designed for modeling SoS architecture with focus on specifying emergent behaviors. SoS architecture is hieratically modeled at abstract level using format and content semantics structural representation, further it uses rewriting logic for describing and reasoning system behavior.

*3) SoS Architectural Frameworks:* SoS Architecture Frameworks (AF) are an efficient way to present different viewpoints [36], [37] in relation to architecture description with separation of concerns for various stakeholders based on well-defined software architecture practices [50]. SoS architecture is designed using UPDM (Unified profiles for DODAF[5] and MODAF[6] with UML/SysML tool support to describe dynamic models [38]. In an effort to design complex systems Daro Delgado et al. [39] proposed a model for executable architecture for SoS using DODAF 2.0. Authors used DODAF standard artifacts on associated data using Petri Nets standard notion to validate and test various elements of architecture from dynamic aspects of SoS, paving the way towards architecture analysis. SergioLuna et al. [40] proposed a framework to model the outcomes of possible interactions of CSs. This approach is a mix of various frameworks of DODAF and Design Structure Matrix (DSM) to efficiently describe the flow of information about constituent systems. SOSA [41] is a SoS AF, specially designed for production industries. Production scenarios for SoS are described and the system is modeled at the structural level.

## B. Ontologies to model SoS

Semantic Ontologies based on languages like OWL, RDF, and RDFS [42] have attained importance in the Software Engineering domain [43] for modeling complex systems. A formal Ontology for architecture descriptions was proposed by Milena Guessi et al. [44] named as ontolAD based on OWL2 for modeling SoS architecture but it lacks essential capability to reason about dynamic architecture. Henrie [45] revisited common terms used for traditional systems and extended them to bring under a framework where systems engineers could use those terms for better understanding with the same vocabulary to solve complex design of SoS. Moschoglou et al. [46] proposed a semantic ontology framework for modeling the architecture of federated SoS for creating service-oriented ubiquitous systems. This framework is based on WSMO (Web Service Modeling Ontology) using semantic annotations for web services. However, it does not describe architectural elements communication protocols and their possible configurations at design time to be realized at runtime.

---

[5]https://dodcio.defense.gov/Library/DoD-Architecture-Framework/dodaf20*viewpoints*/
[6]https://www.gov.uk/guidance/mod-architecture-framework

TABLE II
ANALYSIS OF SEMI-FORMAL SoS ARCHITECTURE MODELING APPROACHES

| Semi-formal SysML Profiles | Structural Description | Behavioral Description | Model unknown Dynamic Reconfigurations | Analysis support with multi-viewpoints | Ability to Model Quality Attributes |
|---|---|---|---|---|---|
| AMADEOS [31, 33] | +++ | +++ | – | ++ | – |
| SysADL [34] | +++ | +++ | – | ++ | – |
| OCL+SysML extensions [35] | +++ | +++ | – | +++ | ++ |
| DANSE [55] | +++ | +++ | – | ++ | ++ |
| *Legends*: +++ means: Fully Supported, ++ means: Partially Supported, – means: Not Supported at all ||||||

## IV. ISSUES ON EXISTING MODELING APPROACHES

Through a detailed critical analysis of architectural approaches described in the section above, we synthesize their abilities with respect to formalism they offer, and to extent they model stochastic SoS dynamic architecture to conform system quality attributes.

SysML brings lots of benefits for modeling SoS architecture, especially it utilizes MDE toolset for models transformation with well-defined profiles at structural and behavioral levels, but lacks essential capability to reason about stochastic nature of SoS. DoDAF architectural specifications show some promise for describing quality attributes of complex system however, it is not able to describe, reason and forecast about dynamic aspects of SoS.

Similarly formally founded ADLs (CML and SosADL) have syntax and semantics for designing SoS at the structural and behavioral level. However, none of these ADLs with underlying formalism of CSP for CML and CCS for SosADL have considered the integration of probabilistic modeling and reasoning capabilities to manage compositions at runtime without compromising runtime quality attributes. Moreover, these ADLs are unable to describe SoS CSs and connectors in order to be able to reason and forecast about runtime changes to unplanned interventions.

An SoS-ADL should not only be able to model structure and behavior of the system but also should have the ability to describe and analyze quality attributes like performance, security and reliability and this provides strong evidence to enrich existing ADLs with MDE methodology to develop the framework for modeling dynamic architecture in order have best alternative architectures and make timely architectural design decisions.

Some of prominent issues related to these SoS modeling techniques identified are following:

- Most of the semi-formal approaches based on SysML coupled with profiles are able to describe structural architecture but are unable to model and reason about dynamicity of SoS essentially due to the inability of underlying notations to deal with complexity of runtime architecture.
- Formally founded ADLs (CML & SosADL) are not fully capable to deal with non-deterministic nature of SoS architecture to forecast stochastic SoS architectural states and their impact on SoS quality attributes.
- Existing ADLs and frameworks are not enriched with syntax and semantics in order to encapsulate time-based changes to be able to model and reason unknown configurations of SoS to guarantee the system requirements conformance with adverse changes.
- There is no probability distribution based mechanism that allows stochastics SoS models to evolve from a runtime perspective over a period of time to view concrete configurations states qualitatively and quantitatively.
- Existing techniques lack an integrated approach for modeling SoS architectures where models need to be simulated for verification and validation that allow to perform analysis to see the impact of dynamic reconfiguration on system quality attributes.

## V. FUTURE RESEARCH DIRECTIONS

In this section, we present future research directions and guidelines for modeling SoS architecture to deal with the research issues identified in previous section. SoS Dynamic reconfiguration is an NP-complete problem that requires, rigorous and well established formal approaches to deal with unknown configurations. Instead of developing altogether a new ADL, it is rather feasible and cost-effective to extend existing SoS-ADLs. Such an extended SoS-ADL integrated with Model-based framework shall be able to model and validate correct and consistent SoS architectures. This will allow various stakeholders to evaluate SoS models and make better design decisions. For this CML/SosADL formally, based on process algebras promise to be more suitable languages to be extended and integrated. We outline following research directions and steps to overcome existing limitations:

TABLE III
SUMMARIZED ANALYSIS OF FORMAL SoS- ADLs

| Formal ADLs for SoS | Formalism support | Structural Description | Behavioral Description | Model unknown Dynamic Reconfigurations | Analysis support with multi-viewpoints | Model Quality Attributes |
|---|---|---|---|---|---|---|
| CML | CSP | +++ | +++ | – | ++ | – |
| SosADL | Pi-calculus (CCS,CCP) | +++ | +++ | – | ++ | – |
| INSAC(IPL+IML) | Pi-calculus | +++ | +++ | – | ++ | ++ |
| ArchSoS | N/A | +++ | ++ | – | – | – |
| **Legends**: +++ means: Fully Supported, ++ means: Partially Supported, – means: Not Supported at all ||||||||

### A. Need for modeling Stochastic SoS

To deal with unknown configurations, Stochastic Process Algebra (SPA) [47], [48] offers certain syntax and semantics to specify the probabilistic distribution over a period of time while integrating quantitative and qualitative modeling of complex systems. Due to its underlying characteristics of compositionality, domain knowledge encapsulation, inference of quantitative measures like performance and the ability to predict non-deterministic behaviors SPA has been widely used to model distributed complex systems. In relation to SoS characteristics, SPA in combination with traditional process, algebra promises to be suitable to model such stochastic structural changes. Existing ADLs for SoS are required to be enriched with probabilistic modeling capabilities to model stochastic behaviors with actions and time-based transitions providing qualitative measures of the emerging configurations.

### B. Need to enrich existing Syntax and Semantics with SPA

Through algebraic process approach, we can only determine the behavior of the interacting processes i.e. CSs forming an emergent behavior but we cannot predict the probability of occurring particular configuration. Existing formal SoS- ADLs need to enrich syntax and semantics to deal with dynamic reconfigurations by incorporating stochastic operators and semantics. For this existing formalism of CML/SosADL needs to be extended with SPA stochastic process algebra in order to deal with non-deterministic time-based variations in architecture structure. A CSP/CCS model consists of the process as $P$ that engages in some action as $\alpha$ forming a CSP model as:

$M \Rightarrow \alpha.P$

$M \Rightarrow Semantic\ Ops. \Rightarrow LTS$

The semantics of formal model are described with Labeled Transition System (LTS) also called axioms for particular actions of the processes. Transition rules are subsequently used for forming architectural behaviors. SoS CSs internal behaviors need not to be known to the external world and it only needs formal semantics based on transition axioms to present its external behavior with certain constraints and communicate with other CSs for establishing relationships leading to configurations. SPA shall allow CSP/CCS to incorporate time-based activity $r$ along with actions between the processes as:

$M \Rightarrow (r.\alpha)P \Rightarrow Semantic\ Ops. \Rightarrow Transitions$

$M \Rightarrow M\text{-}LTS \Rightarrow CTMC$

By applying SPA to CSP/CCS with semantic operations, M-Labelled Transition Systems (M-LTS) will be able to evolve further generating Continuous Time Markov Chains[7] (CTMC) with probabilistic information to transition from current state of the system to next state. With the incorporation of Markov Chains, system architects can not only predict the possible configurations but can measure the runtime quality attributes of SoS quantitatively.

### C. Extend SoS-ADL

Once the underlying formalism is extended of targeted process algebra, grammar for the selected ADL can be integrated by incorporating MDE technology infrastructure. Eclipse Modeling Framework (EMF) provides Ecore for models transformations and code generations allowing to generate models, transform and perform suitable analysis. Meta-models could be defined to form DSLs to describe architectural concepts in a target language by enhancing syntax and semantics. Domain Ontologies for SoS architecture should be defined to capture structural and behavioral concepts related to SoS system along with quality attributes with underlying mediated constraints specifications. Semantics and reasoning should be incorporated into ADL using suitable logic techniques to describe SoS structural changes. The extended ADL should allow to specify certain constraints to adopt to particular configurations with their quality attributes.

### D. Need of a Framework for modeling and Simulation

A well-formed integrated framework is required, to model, reason and validate SoS architecture. The future framework should not only allow the system architects to describe and reason about the architecture at an abstract level but it should have the capability to further analyze possible configurations, select valid architectures for better results to develop resilient and fault tolerant high-quality systems. For this, SoS architecture model with realistic representations of possible configurations needs to be modeled and simulated with model-based stochastic formalism. From abstract architectural model, Possible stochastic configurations states should be analyzed with respect to time along with the ability to describe quality attributes quantitatively for each state.

---

[7]In System modeling perspective Markov Chains can be described as transition matrix exhibiting system evolution from state to state with a time-based function of probabilistic distribution. Markov Chains can be continuous or discrete depending on the system states changes with time variation.

Further validation of the models should be performed with statistical model checking and simulation techniques.

## VI. CONCLUSIONS

In this research paper, we have explored SoS architecture modeling approaches by developing a taxonomy to assess their ability to specify, reason and evolve SoS architecture especially to manage dynamic structural changes. We emphasized the importance of SoS dynamic architecture modeling by describing key concepts, possible structural changes and their impact on SoS architecture. This research reveals that existing syntax and semantics of the ADLs are not capable to forecast the non-deterministic structural changes due to the underlying characteristics of SoS. Based on identified gaps, we propose future extensions in existing formalism of SoS-ADLs woven into model-based architecture framework to model and validate dynamic SoS architectures. The ADL extensions should be woven into a model-based framework by incorporating stochastic modeling, model checking and simulation capabilities such as CTMC to predict possible SoS states and manage quality attributes quantitatively. This will allow system architects to make better architecture design decisions by selecting appropriate SoS architecture models that conform to system properties.